# Energy Transport Property of Charged Particles with Time-Dependent Damping Force via Manifold-Based Analysis Approach

Hao Zhang[*], Pengcheng Luo, Huifang Ding

*State Key Lab of Electrical Insulation & Power Equipment, School of Electrical Engineering, Xi'an Jiaotong University, Xi'an 710049*

[*]haozhang@xjtu.edu.cn

**Abstract** This paper deals with the energy transport properties of charged particles with time-dependent damping force. Based on the proposed nonlinear dimensionless mapping, the stability and dynamical evolution of the particle system is analyzed with the help of manifold-based analysis approach. It has been found that the particle system possesses two types of energy asymptotic behaviors. More significantly, the underlying mechanism of an "energy barrier" is uncovered, *i.e.*, one generalized invariant spanning curve emerges in the dissipative particle system. These results will be useful to enrich the energy transport behavior knowledge of the particle system.

## 1. Introduction

Fermi acceleration (FA), a strange physical phenomenon where a particle acquires unlimited energy, has given an adequate explanation for the origin of high energy of cosmic rays [1]. Since then, some investigation has been carried out on the dynamical characteristics of various particle systems in the field of physics [2-7]. It has been found that many particle systems possess a very complex mixed phase space structure including chaotic sea, Kolmogorov-Arnol'd-Moser (KAM) islands and a set of invariant spanning curves [8-12]. In actual applications, particle systems can be inevitably subjected to the dissipative forces resulting from inelastic collision effect [13-15]and damping effect [16-19], and particularly these dissipative forces have significant effect on the energy transport property of particles. Since the inelastic collision effect is sustained only in the instant of collision while the damping effect does ceaselessly in the process of movement, the damping effect appears to be more interesting for the purpose of enriching the entire dynamical knowledge of the particle systems.

In this paper, the energy transport property of the charged particle system with time-dependent damping force will be investigated. Based on the nonlinear dimensionless mapping, the stability and dynamical evolution will be analyzed with the help of eigenvalue analysis method and invariant manifold analysis in phase space. In addition, the energy transport behavior will be evidently classified by the basin of attraction. Finally, theoretical analysis will be verified by numerical results.

## 2. System Description and Modeling

In this paper, we will investigate the energy transport property of the charged particle system with time-dependent damping force. The system consists of a metal particle of mass $m$ or an ensemble of non-interacting particles moving between two metallic blocks. One block is fixed at $x = l$, while the other is moving in the form of the equation $x = \delta \cos(\omega t + \phi_0)$. Since the two blocks are charged with opposite charges, an electric field $E$ can be generated. So the particle will suffer from an electric field force $F_e = qE$ as long as it moves between the two blocks [8].

Provided that the particle moves in such a specific medium as air, water and oil, then the movement of the charged particle will be affected by the viscous drag force. In nature, the drag force is one class of typical time-dependent damping forces. Here the drag force's magnitude is regarded as proportional to the particle's velocity, *i.e.*, $F_d = \eta' v$. (where $\eta'$ is the viscosity). So the resultant force $F = F_e + F_d = Eq - \eta' v$ acts along the entire trajectory of the charged particle. Based on Newton's second law, we have

$$m \, dv / dt = Eq - \eta' v \tag{1}$$

Supposing that the initial condition of the particle is $v = v_n$ at the instant $t = t_n$

with the initial position at $x = 0$, then by solving the differential equation (6), we obtain the velocity of the particle

$$v(t) = (v_n - qE/\eta')e^{-\eta't/m} + qE/\eta' \qquad (2)$$

The corresponding position is derived by a direct integration of Eq.(2)

$$x(t) = \frac{v_n - qE/\eta'}{\eta'/m}\left(1 - e^{-\eta't/m}\right) + (qE/\eta')t \qquad (3)$$

The mapping $T_d^{'}$ is derived by a discrete sampling of $v(t)$ at the instant of collisions with the moving block.

$$T_d^{'}:\begin{cases} v_{n+1} = \left|(v_n - qE/\eta')e^{-\eta'\Delta t_{n+1}/m} + qE/\eta' - 2\omega\grave{o}\sin(\omega t_{n+1} + \phi_0)\right| \\ t_{n+1} = t_n + \Delta t_{n+1} \end{cases} \qquad (4)$$

where the term $-2\omega\grave{o}\sin(\omega t_{n+1} + \phi_0)$ gives the corresponding fraction of the particle's velocity gained or lost in each collision with the moving block, and the time interval $\Delta t_{n+1}$ between two collisions with the oscillating block must satisfy

$$\frac{v_n - qE/\eta'}{\eta'/m}\left(1 - e^{-\eta'\Delta t_{n+1}/m}\right) + (qE/\eta')\Delta t_{n+1} = 2l \qquad (5)$$

For simplicity, a new set of dimensionless variables $V = v/(\omega l)$, $\delta_1 = \eta'/(\omega l)$, $\delta_2 = qE/(m\omega^2 l)$ and $\varepsilon = \grave{o}/l$ are defined. Here let $\phi_{n+1} = \omega t_{n+1} + \phi_0$, then Eq.(4) can be rewritten as the following nonlinear mapping

$$T_d:\begin{cases} V_{n+1} = \left|V_n - 2\delta_1 + \delta_2\Delta\phi_{n+1} - 2\varepsilon\sin\phi_{n+1}\right| \\ \phi_{n+1} = (\phi_n + \Delta\phi_{n+1})\,\mathrm{mod}\,2\pi \end{cases} \qquad (6)$$

where $\Delta\phi_{n+1}$ meets the following condition

$$(\frac{V_n}{\delta_1} - \frac{\delta_2}{\delta_1^2})(1 - e^{-\delta_1\Delta\phi_{n+1}}) + \frac{\delta_2}{\delta_1}\Delta\phi_{n+1} = 2 \qquad (7)$$

The fixed points of mapping (6) can be obtained by calculating $V_{n+1} = V_n = V^*$ and $\phi_{n+1} = \phi_n = \phi^*$

$$\begin{cases} \phi^* = \arcsin((k\delta_2\pi - \delta_1)/\varepsilon) \\ V^* = (2\delta_1 - 2k\delta_2\pi)/(1 - e^{-2k\delta_1\pi}) + \delta_2/\delta_1 \end{cases} \quad (8)$$

where $k$ is a positive integer, the parameters $\delta_1$, $\delta_2$ and $\varepsilon$ should satisfy Eq.(9)

$$\left| (k\delta_2\pi - \delta_1)/\varepsilon \right| \le 1 \quad (9)$$

From Eq.(8), it can be clear that there are $k$ pairs of fixed points in this particle system. For brief, two symbols $F_1$ and $F_2$ denote one pair of fixed points.

## 3. Numerical Results and Manifold-Based Analysis

Since the particle system can exhibit a great variety of energy transport behaviors such as energy diffusion or decay in the process of system dynamical evolution, the energy transport property of charged particle is tightly relevant to the dynamical one. So in this section, the energy transport behaviors will be investigated from the standpoint of dynamics.

### 3.1. a Glimpse at Energy Transport Behaviors via Numerical Simulation

When $\delta_1 = 1.01 \times 10^{-3}$, $\delta_2 = 2 \times 10^{-4}$ and $\varepsilon = 1 \times 10^{-3}$, Fig.1 shows the time-domain waveforms of the velocity $v$ from different initial conditions. From Fig.1(a), it is found that when the initial conditions are (0.4, 1.7) and (0.13, 3), the particles eventually reach two different steady states, *i.e.*, fixed points corresponding to $k = 1$ and $k = 2$, respectively. However, when initial conditions are (0.4, 0) and (0.13, 0), the interesting thing is that there is another steady state fluctuating about $V \approx 0.198$ (seen in Fig.1(b)). This means that a strange energy transport behavior occurs in the particle system.

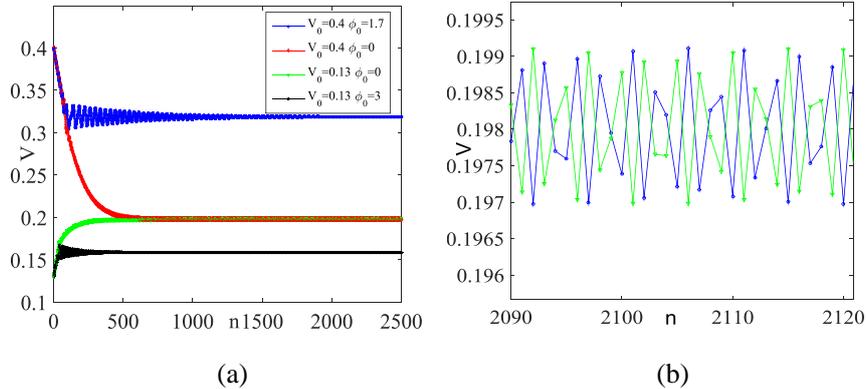

(a)                    (b)

Fig.1 (a) Time waveforms of velocity $V$ from different initial conditions (b) Close-up view

## 3.2. Manifold-Based Analysis in Phase Pace

Based on the nonlinear mapping (6), the above transport behaviors will be investigated in terms of invariant manifolds in phase space. To start with, we began to investigate the local stability of the fixed points, which is determined by the eigenvalues of the Jacobian matrix in the neighborhood of the fixed point.

The Jacobian matrix $J$ in the neighborhood of fixed points $(\phi^*, V^*)$ is derived as

$$J = \begin{bmatrix} \dfrac{\partial V_{n+1}}{\partial V_n} & \dfrac{\partial V_{n+1}}{\partial \phi_n} \\ \dfrac{\partial \phi_{n+1}}{\partial V_n} & \dfrac{\partial \phi_{n+1}}{\partial \phi_n} \end{bmatrix}_{(\phi^*, V^*)} \tag{10}$$

Then the eigenvalue $\lambda$ can be derived by solving $\det(\lambda I - J) = 0$, where $I$ is unit matrix.

When $\delta_1 = 1.01 \times 10^{-3}$ and $\varepsilon = 1 \times 10^{-3}$, Fig.2 gives the eigenvalue loci of Jacobian matrix $J$ in the neighborhood of the fixed points $F_1$ and $F_2$ with the increase of $\delta_2$, respectively. From Fig.2(a), one can see that since $\lambda_1 > 1$ and $\lambda_2 < 1$ always hold under the constrained condition of Eq.(9), the fixed points $F_1$ are saddle points. This implies that the particle motion on these points is unstable. From Fig.2(b), it can be seen that the modulus of a pair of conjugate complex eigenvalues $\lambda_{1,2}$ are smaller than 1 at all times. Accordingly, the fixed points $F_2$ are attractor points. This indicates the motions of particle are always stable on the points $F_2$.

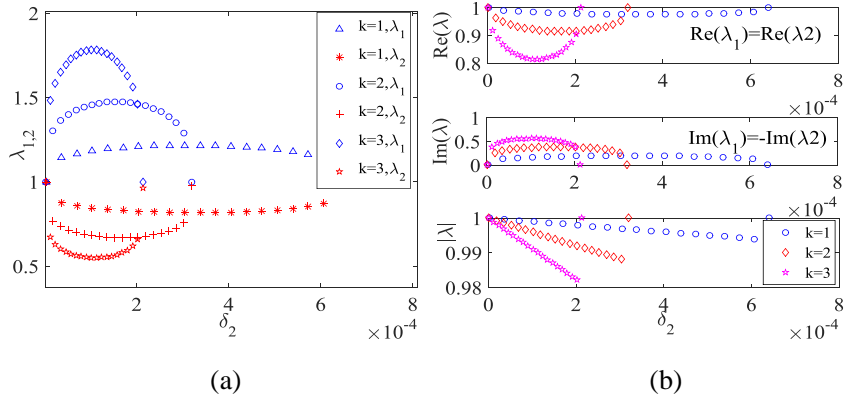

Fig.2 Eigenvalue loci of Jacobian $J$ in the neighborhood of (a) fixed points $F_1$ (b) fixed points $F_2$ with the increase of $\delta_2$

When the inequality (9) is satisfied, the largest integer $k$ is $k = 3$. This means that there are three pairs of fixed points, where one is saddle point and the other is attractor point for each $k$. Here we assume that $S_k$ and $A_k$ denote a saddle point and a attractor point, respectively, then $S_{k1}$, $S_{k2}$ and $U_{k1}$, $U_{k2}$ are two stable manifolds and two unstable manifolds corresponding to the saddle point $S_k$, respectively.

Fig.3(a) shows the stable and unstable manifolds corresponding to all three saddle points $S_1$, $S_2$ and $S_3$. Fig.3(b) gives a close-up view near $V \approx 0.198$ of Fig.3(a). It is clear that the manifolds of saddle points $S_1$ and $S_2$, $S_3$ are completely separated by an "energy barrier". To further analyze the energy evolution of the particles, Fig.3(c), (d) and (e) illustrate the manifolds near the fixed points $S_k$ and $A_k$, respectively. As shown in Fig.3(c), the two stable manifolds $S_{11}$ and $S_{12}$ come from the high energy region of phase space. The unstable manifolds $U_{11}$ evolves to the attractor point $A_1$, while $U_{12}$ evolves to the low energy region but never passes through the " barrier". From Fig.3(d) and (e), one can see that unlike $S_1$, the stable manifolds $S_{21}$, $S_{22}$ and $S_{31}$, $S_{32}$ origin from the low energy region. In addition, the unstable manifolds $U_{22}$ and $U_{32}$ evolve towards the attractor points $A_2$ and $A_3$, respectively. In comparison with $U_{22}$, $U_{21}$ evolves to the higher energy but never passes through the same "barrier". Note that some points of $U_{31}$ evolve to the same region as $U_{11}$ and $U_{21}$ while some of them converge to the attractor point $A_2$. Fig.3 allows us to conclude that the particle system can exhibit two types of energy asymptotic behaviors, *i.e.*, the attractor points $A_1$, $A_2$ and $A_3$, and the " energy barrier".

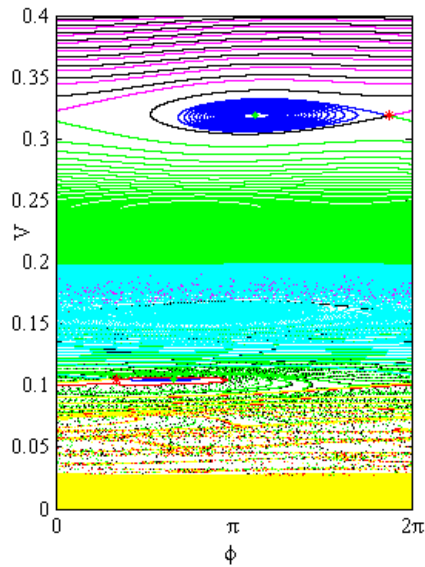

(a)

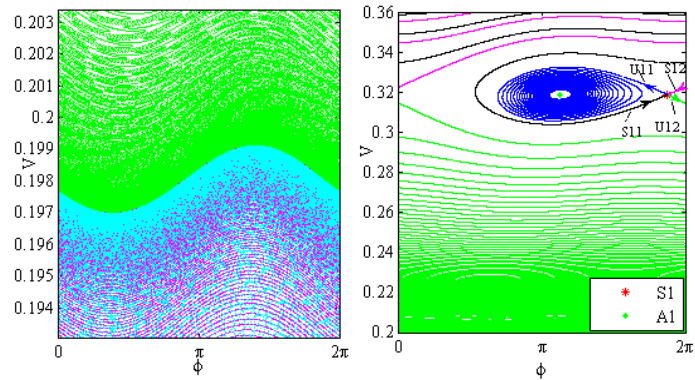

(b)                              (c)

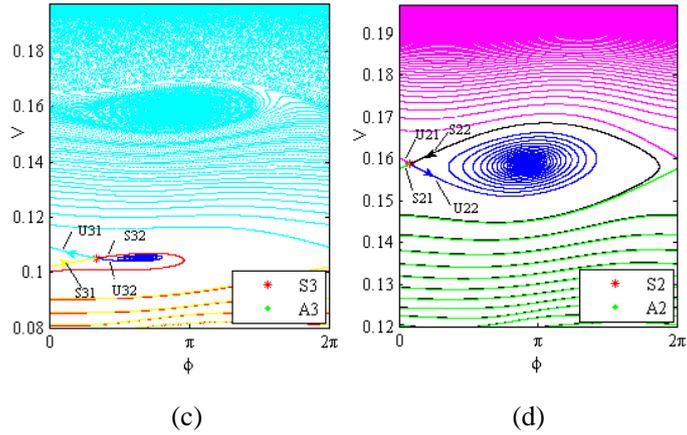

(c)                  (d)

Fig.3 (a) Stable and unstable manifolds of all saddle points (b) Close-up view near $V \approx 0.198$
(c), (d) and (e) Manifolds near the fixed points for $k = 1$, $k = 2$ and $k = 3$, respectively

In order to make a clear classification of the energy transport behaviors, one of the most powerful tools characterizing the asymptotic behavior of a dynamical evolution known as basin of attraction is presented in Fig.4, where the red, light blue and blue region represent the basin of attraction corresponding to $A_1$, $A_2$ and $A_3$, respectively. If compared, most of the region is covered by the mint green, which represents the steady state of these initial conditions is the "energy barrier". Thus, it is concluded that the asymptotic behavior to the "barrier" is a special but global energy transport behavior in this particle system. Besides the initial conditions converging to the attractor points $A_1$, $A_2$, $A_3$, most of particles reach the "barrier" as shown in Fig.4.

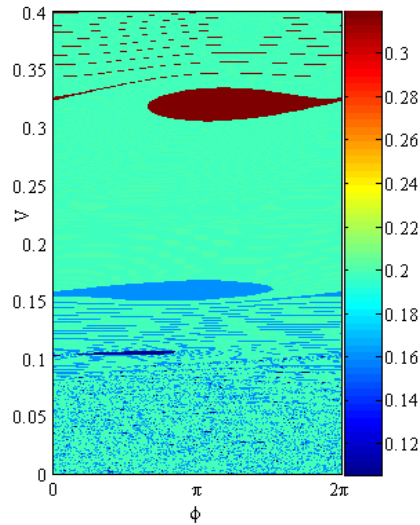

Fig.4 Basin of attraction corresponding to $A_1$, $A_2$, $A_3$ and the "barrier"

From the numerical results, it is shown that a special but global energy transport behavior to the "barrier" occurs in the particle system. To give a more evident explanation, we choose the initial condition (0.4,0) so that the steady state is the "barrier". From the phase portrait in Fig.5, we can observe that the "barrier" surprisingly appears in the form of one invariant spanning curve because such invariant curve usually emerges in non-dissipative particle system [20, 21]. Now the following question may be posed: whether dose the "barrier" contribute to the fact that the resultant force $F$ between the two blocks equals to zero?

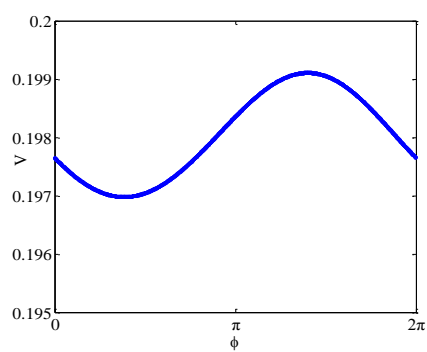

Fig.5 Phase portrait

If the resultant force $F$ is zero, *i.e.*, $Eq - \eta' v = 0$, then the variable $\bar{v}$ will be used to express the equilibrium position in Fig.5

$$\bar{V} = \delta_2 / \delta_1 \tag{11}$$

Obviously, the position of the "barrier" is determined by the parameters $\delta_1$ and $\delta_2$. To make a verification of Eq.(11), Fig.6 gives the numerical and the theoretical results of the relationship between the variable $\bar{v}$ and $\delta_2$ when $\delta_1$ is fixed at $1.01 \times 10^{-3}$. It is found that the two results are in good agreement.

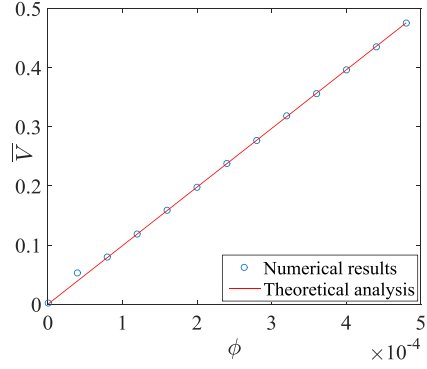

Fig.6 Position of the "barrier" as a function of $\delta_2$

Based on the above analysis, it is found that the "energy barrier" in this particle system is essentially an invariant spanning curve. Different from non-dissipative particle system, the position of the invariant spanning curve depends on the system parameters rather than the initial velocity. In non-dissipative particle system, the energy of the particle on this invariant curve will get back to the initial energy again after some collisions. On the contrary, when the particle system suffers from time-dependent damping force, the initial velocity will begin increase or decrease to the invariant curve, and eventually remain on the invariant curve, as shown in Fig.1.

## 4. Conclusions

In summary, we have investigated the energy transport properties of charged particles with time-dependent damping force. It has been found from the manifold-based analysis that there are two types of energy transport behaviors in this particle system. Surprisingly, the charged particle system possesses one generalized invariant spanning curve, an "energy barrier", whose position in phase space is directly determined by the system parameters $\delta_1$ and $\delta_2$, *i.e.* the electric field force $F_e$ and the drag force $F_d$. These results can be applied to such charged particle systems suffering from a resultant force $F = qE - \eta'v^{\gamma}$, where $\eta'$ is viscosity, depending on the specific medium.

## Acknowledgements

This work was supported by National Natural Science Foundation of China (Grant

No. 51577141).

# References

[1]     S. Bhaduri and D. Ghosh, "Study of Void Probability Scaling of Singly Charged Particles Produced in Ultrarelativistic Nuclear Collision in Fractal Scenario," *Advances in High Energy Physics,* 2016.

[2]     K. Kotera and A. V. Olinto, "The Astrophysics of Ultrahigh-Energy Cosmic Rays," *Annual Review of Astronomy and Astrophysics, Vol 49,* vol. 49, pp. 119-153, 2011.

[3]     P. Blasi, "The origin of galactic cosmic rays," *Astronomy and Astrophysics Review,* vol. 21, p. 00159, Nov 5 2013.

[4]     A. Veltri and V. Carbone, "Radiative intermittent events during Fermi's stochastic acceleration," *Physical Review Letters,* vol. 92, p. 143901, Apr 9 2004.

[5]     K. Kobayakawa, Y. S. Honda, and T. Samura, "Acceleration by oblique shocks at supernova remnants and cosmic ray spectra around the knee region," *Physical Review D,* vol. 66, p. 083004, Oct 15 2002.

[6]     T. Kroetz, A. L. P. Livorati, E. D. Leonel, and I. L. Caldas, "Global ballistic acceleration in a bouncing-ball model," *Physical Review E,* vol. 92, p. 012905, Jul 6 2015.

[7]     T. Ghaffary, "Charged Particles Multiplicity and Scaling Violation of Fragmentation Functions in Electron-Positron Annihilation," *Advances in High Energy Physics,* 2016.

[8]     B. He, H. F. Ding, H. Zhang, and Y. P. Meng, "Nonlinear dynamics of charged particle slipping on rough surface with periodic force," *Nonlinear Dynamics,* vol. 85, pp. 2247-2259, Sep 2016.

[9]     E. D. Leonel, J. K. L. da Silva, and S. O. Kamphorst, "On the dynamical properties of a Fermi accelerator model," *Physica a-Statistical Mechanics and Its Applications,* vol. 331, pp. 435-447, Jan 15 2004.

[10]    B. Liebchen, R. Buchner, C. Petri, F. K. Diakonos, F. Lenz, and P. Schmelcher, "Phase space interpretation of exponential Fermi acceleration," *New Journal of Physics,* vol. 13, p. 093039, Sep 29 2011.

[11]    A. L. P. Livorati, C. P. Dettmann, I. L. Caldas, and E. D. Leonel, "On the statistical and transport properties of a non-dissipative Fermi-Ulam model," *Chaos,* vol. 25, p. 103107, Oct 2015.

[12]    E. V. B. Leite, H. Belich, and K. Bakke, "Aharonov-Bohm Effect for Bound States on the Confinement of a Relativistic Scalar Particle to a Coulomb-Type

Potential in Kaluza-Klein Theory," *Advances in High Energy Physics,* 2015.

[13] E. D. Leonel and R. E. de Carvalho, "A family of crisis in a dissipative Fermi accelerator model," *Physics Letters A,* vol. 364, pp. 475-479, May 14 2007.

[14] A. L. P. Livorati, I. L. Caldas, C. P. Dettmann, and E. D. Leonel, "Crises in a dissipative bouncing ball model," *Physics Letters A,* vol. 379, pp. 2830-2838, Nov 6 2015.

[15] A. A. Deriglazov and W. G. Ramirez, "Ultrarelativistic Spinning Particle and a Rotating Body in External Fields," *Advances in High Energy Physics,* 2016.

[16] D. G. Ladeira and E. D. Leonel, "Dynamics of a charged particle in a dissipative Fermi-Ulam model," *Communications in Nonlinear Science and Numerical Simulation,* vol. 20, pp. 546-558, Feb 2015.

[17] E. D. Leonel and P. V. E. McClintock, "Effect of a frictional force on the Fermi-Ulam model," *Journal of Physics a-Mathematical and General,* vol. 39, pp. 11399-11415, Sep 15 2006.

[18] D. F. Tavares, E. D. Leonel, and R. N. Costa, "Non-uniform drag force on the Fermi accelerator model," *Physica a-Statistical Mechanics and Its Applications,* vol. 391, pp. 5366-5374, Nov 15 2012.

[19] D. F. M. Oliveira and M. Robnik, "In-flight dissipation as a mechanism to suppress Fermi acceleration," *Physical Review E,* vol. 83, p. 026202, Feb 14 2011.

[20] E. Fermi, "On the origin of the cosmic radiation," *Physical Review,* vol. 75, pp. 1169-1174, 1949.

[21] E. D. Leonel, J. K. L. da Silva, and S. O. Kamphorst, "On the dynamical properties of a Fermi accelerator model," *Physica a-Statistical Mechanics and Its Applications,* vol. 331, pp. 435-447, Jan 15 2004.